\documentstyle[12pt,epsf]{article}

\def\lesssim{\mathrel{\mathpalette\vereq<}}

\makeatletter
\def\vereq#1#2{\lower3pt\vbox{\baselineskip1.5pt \lineskip1.5pt
\ialign{$\m@th#1\hfill##\hfil$\crcr#2\crcr\sim\crcr}}}
\makeatother

\begin{document}

\begin{titlepage}
\begin{center}
\today     \hfill    LBNL-4-0077 \\
~{} \hfill UCB-PTH-97/01  \\
~{} \hfill hep-ph/9703259\\

\vskip .1in

{\large \bf Can The Supersymmetric Flavor Problem Decouple?}\footnote{
This work was supported in part by  
the U.S. Department of Energy under Contracts DE-AC03-76SF00098 and 
DE-FG-0290ER40542 and in part by the National Science Foundation under 
grant PHY-95-14797.  NAH was also supported by NSERC, and HM by the 
Alfred P. Sloan Foundation.}

\vskip 0.3in

Nima Arkani-Hamed and Hitoshi Murayama

\vskip 0.05in

{\em Theoretical Physics Group\\
     Earnest Orlando Lawrence Berkeley National Laboratory\\
     University of California, Berkeley, California 94720}

\vskip 0.05in
{\rm and}
\vskip 0.05in

{\em Department of Physics\\
     University of California, Berkeley, California 94720}

\end{center}

\vskip .1in

\begin{abstract}
It has been argued that the squarks and sleptons of the first and 
second generations can be relatively heavy without destabilizing the 
weak scale, thereby improving the situation with too-large 
flavor-changing neutral current (FCNC) and CP violating 
processes. In theories where
the soft supersymmetry breaking parameters are generated at a
high scale (such as the Planck scale), we show that such a mass 
spectrum tends to drive the scalar top mass squared $m_{\tilde{Q}_3}^2$
negative from 
two-loop renormalization group evolution.  
Even ignoring CP violation and allowing $O(\lambda) \sim .22$ alignment,
the first two generation
scalars must be heavier than 22~TeV to suppress FCNC. 
This in turn requires the boundary 
condition on $m_{\tilde{Q}_3} > 4$~TeV to avoid negative $m_{\tilde{Q}_3}^2$ 
at the weak scale.
Some of the models in the literature employing the anomalous U(1) in string 
theory are excluded by our analysis.
\end{abstract}

\end{titlepage}

\newpage

The biggest embarassment of low-energy supersymmetry (SUSY)
is the 
flavor problem: the superparticles may generate too-large
flavor-changing (FC) effects  
such as $K^{0}$-$\bar{K}^{0}$ mixing or $\mu \rightarrow e\gamma$, or 
too-large neutron and electron electric dipole moments.  
Traditionally, one assumes that the SUSY breaking parameters 
are universal and real at a high scale to avoid these problems, as  
done in the ``minimal supergravity'' framework.  Supergravity, 
however, does not have a fundamental principle to guarantee the 
universality of scalar masses nor their reality.  Several ideas have 
been proposed to solve this supersymmetric flavor problem.  If 
SUSY breaking is mediated by gauge interactions \cite{GM}, or 
if the dominant SUSY breaking effect is in the dilaton 
multiplet of string theory \cite{dilaton}, the soft breaking parameters 
are flavor blind and the problem is eradicated.  Alternately, flavor 
symmetries can guarantee sufficient degeneracy amongst the first- and 
second-generation sfermions \cite{nonabelian}, or alignment between 
quark and squark mass matrices \cite{NS}.

It would be simplest, however, to push up the masses of the first and
second generation scalars high enough to avoid
the flavor problem \cite{DKS,DG,PT,CKN}.  Since the first two generations have 
small Yukawa couplings to the Higgs doublets, it is conceivable 
that they can be heavy while maintaining natural electroweak symmetry 
breaking (EWSB), which is the very motivation for low-energy 
SUSY.  We of course still need to keep the masses of third 
generation scalars, gauginos and higgsinos close to the weak 
scale for this purpose.

Such a spectrum was studied in \cite{DG} and it was argued that  
too-heavy first- and second-generation scalars lead to a fine-tuning 
in EWSB because they give a too large  
contribution to the Higgs mass squared via two-loop 
renormalization group equations (RGE).  It was concluded that the heavy
scalars need to  be (at least) lighter than 5~TeV to avoid a fine-tuning 
of more than 10~\% in EWSB.
A subsequent analysis \cite{PT} 
required that the mass splitting between the different generations be
preserved by the two-loop RGEs, obtaining a similar constraint.  
However, the constraints based on this type of discussion are
somewhat subjective: the results depend on how large a fine-tuning one allows, 
or exactly what is meant by the preservation of the mass splitting.  More 
recently, such a split mass 
spectrum was argued to be best from the phenomenological point of view
\cite{CKN}.  
Furthermore, it was pointed out that the $D$-term contributions from 
the anomalous U(1) gauge group in string theory may naturally lead to 
such a split mass spectrum \cite{BD,DP}.  Therefore, it is useful to study 
the phenomenological viability of this type of spectrum.

The purpose of this letter is to point out that such a split scalar 
mass spectrum tends to drive the mass squared of third 
generation squarks/sleptons negative, breaking color and charge.   
This constraint is purely phenomenological and does not depend on 
any naturalness criteria.  Indeed, the mass patterns
proposed in some stringy anomalous U(1) models do not satisfy our 
constraint and are hence not phenomenologically viable. Throughout the 
letter
we assume that SUSY breaking parameters are generated at a
high scale such as the Planck scale. Our results are then not an immediate
concern for models where all effects of SUSY breaking shut off 
at scales of a few orders of magnitude above the weak scale, as in the
``more minimal" scenario \cite{CKN}, since the negative
contributions to the scalar masses are not enhanced by a large logarithm.
Nevertheless, our results are strong enough to suggest that 
in any concrete realization of a ``more minimal" scenario,
a detailed check of the radiative corrections to third generation scalar
masses must be done to ensure that they are not driven negative.      

Our analysis has three steps.  We first determine the
minimum 
mass of the first- and second-generation scalars which make the SUSY
contribution to $K^0-\bar{K}^0$ mixing smaller than the observed value. 
Next, we determine the smallest allowed ratio of the scalar mass of
the third generation to that of other generations
consistent with the requirement that $m_{\tilde{Q}_3}^2$
is not driven negative by the two-loop RGE; this constraint is
independent of the discussion of FCNC.  Finally, we combine the 
two analyses to obtain the minimum boundary value for $m_{\tilde{Q}_3}$
consistent both with $K^0-\bar{K}^0$ mixing and with positivity of 
$m_{\tilde{Q}_3}^2$ at the weak scale. 
We find it is difficult to keep the third-generation scalars 
below the TeV scale, even ignoring CP violation and allowing 
$O(\lambda)$ degeneracy or alignment in 
scalar mass matrices of the first two generations. This observation
strengthens 
the case for flavor symmetries or dynamical mechanisms for degeneracy.

\begin{figure}
\centerline{ \epsfxsize = 0.45\textwidth \epsfbox{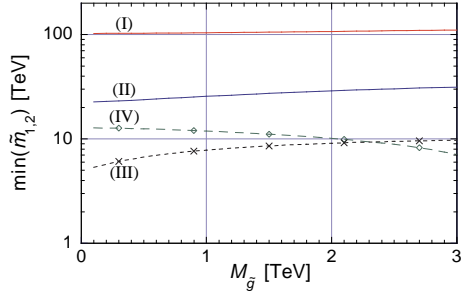} } 
\caption[m1min]{The minimum mass of the first- and second-generation 
scalars $\mbox{min}(\widetilde{m}_{1,2})$ to keep $(\Delta 
m_{K})_{\tilde{q},\tilde{g}}$ $< (\Delta m_{K})_{obs}$ as a function of 
the gluino mass for four different cases: (I) $(\delta_{12}^{d})_{LL} 
= (\delta_{12}^{d})_{RR} = 1$, (II) $(\delta_{12}^{d})_{LL} = 
(\delta_{12}^{d})_{RR} = 0.22$, (III) $(\delta_{12}^{d})_{LL} = 
(\delta_{12}^{d})_{RR} = 0.05$, and (IV) $(\delta_{12}^{d})_{LL}=0.22$ 
and $(\delta_{12}^{d})_{RR} =0$.  Including $O(1)$ CP violating phases 
makes the lower bound stronger by a factor of 13.}
\label{m1min}
\end{figure}

We consider four patterns for the first two generation 
squark mass matrices.  Working in the basis of superfields where the 
down-quark mass matrix is diagonal, it is convenient to characterize the patterns 
by the ratio $(\delta^{d}_{12})$ 
of the off-diagonal $(1,2)$ elements $(m^2_{\tilde{d}})_{12}$ of the 
$\tilde{d}$ mass squared matrices to the average of the squared 
mass eigenvalues $\widetilde{m}^2_{1,2}$, for both left- and right-handed 
scalars. In case (I), the first- and second-generation fields are heavy 
with masses of the same order of magnitude, but with $O(1)$ off-diagonal 
elements, {\it i.e.}\/, $(\delta^{d}_{12})_{LL} = 
(\delta^{d}_{12})_{RR} = 1$.  Case (II) assumes an 
$O(\lambda)$ alignment, $(\delta^{d}_{12})_{LL} = 
(\delta^{d}_{12})_{RR} = 0.22$.
In case (III), we assume that there is some 
small amount of degeneracy $\sim 1/5$ between the first two generation 
scalars, on top of an $O(\lambda)$ alignment, so 
that the off-diagonal elements are $(\delta^{d}_{12})_{LL} = 
(\delta^{d}_{12})_{RR} = 0.05$.  
Finally in case (IV), we assume that the only mixing is between
left-handed squarks and is $O(\lambda)$: $(\delta^{d}_{12})_{LL} = 
0.22$ and $(\delta^{d}_{12})_{RR} = 0$.
This case is motivated by our 
lack of knowledge of the mixing between right-handed quarks, although 
it is somewhat artificial.  Our analysis is then very simple.  We 
require the squark-gluino contribution $(\Delta 
m_{K})_{\tilde{q},\tilde{g}}$ to $K^0$-$\bar{K}^0$ 
mixing (using the formulae in \cite{Masiero}) to be less than the 
observed size $(\Delta m_{K})_{obs}$.
We give the lower bounds on $\widetilde{m}^2_{1,2}$
for each pattern of squark mixing, as a function of the gluino mass 
$M_{\tilde{g}}$.  The results are 
plotted in Fig.~\ref{m1min}.  
In all cases, this lower bound ranges from 100 
TeV to the multi-TeV range.  If one further allows $O(1)$ phases 
in off-diagonal elements, the lower bounds on $\widetilde{m}_{1,2}$ 
become stronger by a factor of 13.
Therefore the scalar masses of the first two generations make 
important {\it negative}\/ contributions 
to the RGE of third-generation scalar masses due to gauge interactions 
at the two-loop level. We thus turn our attention to the RGE analysis.

First note that the heavy first- and second-generation scalars of the
same generation must have 
certain degeneracies among themselves to avoid inducing a too-large 
Fayet--Illiopoulos $D$-term $D_{Y}$ for the hypercharge gauge group at
one-loop.  
Since their mass scale is high, such a contribution would
induce negative mass squared to 
either $\tilde{\tau}$ or $\tilde{L}_{3}$ depending on its sign
\cite{DG,CKN}.  
Therefore we require the scalars within each of the ${\bf 5}^{*}$,
${\bf 10}$ SU(5)-multiplets to be degenerate \cite{foot2}, and
consider cases where $N_5$ of the ${\bf 5}^*$'s and 
$N_{10}$ of the ${\bf 10}$'s are heavy. $N_5 = N_{10} = 2$ is relevant
for all patterns of squark masses (I-IV), 
while $N_5=0,N_{10}=2$ is possible for case (IV).    
Second, we take the 
gaugino masses universal (=$M_0$) at the GUT-scale for simplicity.  
Third, we run all scalar masses starting from the 
GUT-scale $M_{GUT} = 2\times 10^{16}$~GeV.  If the scale where the SUSY 
breaking effects are transmitted is lower, the effects of running will 
be smaller and the constraints will be weaker.  On the other hand, the 
string-derived case starts at the Planck scale and the constraints are 
stronger.  We chose the GUT-scale as 
a compromise for the presentation.
Finally, we  omit all the Yukawa couplings in the RGE: 
since the Yukawa couplings 
always drive the scalar masses smaller, this is a conservative choice.  
Given a  model with specific predictions for the scalar mass 
spectrum, the analysis must be repeated with Yukawa couplings included 
in the RGEs.  Without a concrete model in mind, our choice 
suffices for this letter.

\begin{figure}
\centerline{ \epsfxsize = 0.45\textwidth \epsfbox{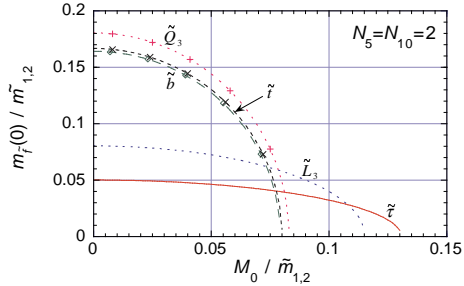} } 
\caption[both]{Constraint on the mass ratio between the 
first- and second-generation scalars $\widetilde{m}_{1,2}$ and the 
third-generation scalars $m_{\tilde{f}}(0)$ from the requirement 
that none of the third-generation scalars acquire negative 
mass squared at the weak scale.  The regions below the curves are
excluded.  Constraints are shown for the case $N_{5} = N_{10} = 2$.  See the 
text for details of our conservative assumptions.}
\label{both}
\end{figure}

We take the two-loop RGEs in the $\overline{\rm DR}'$ scheme \cite{All Star}.
We neglect all two-loop terms subdominant to the
ones involving the heavy scalar masses.  Neglecting Yukawa 
couplings as discussed above, the RGEs have only two important 
contributions: the one-loop gaugino contributions and the two-loop 
contributions from the heavy scalars.  Furthermore, the running of
the heavy scalar masses 
is negligible. The RGE 
for the third-generation scalar species $\tilde{f}$ is then given by
\begin{eqnarray}
\lefteqn{
	\frac{d}{dt} m^{2}_{\tilde{f}} =
		- 8 \sum_{i} \tilde{\alpha}_{i} C_{i}^f M_{i}^{2}
		+ 8 \left[ 
			\left(\frac{1}{2} N_{5} + \frac{3}{2} N_{10} \right)
			\sum_{i} \tilde{\alpha}_{i}^{2} C_{i}^f
			\right. }\nonumber \\
	& & \left.
			+ (N_{5} - N_{10})
			\frac{3}{5} Y_{f} \tilde{\alpha}_{1}
			\left( \frac{4}{3} \tilde{\alpha}_{3}
				- \frac{3}{4} \tilde{\alpha}_{2}
				- \frac{1}{12} \tilde{\alpha}_{1} \right) 
			\right]  \widetilde{m}_{1,2}^{2}.
\end{eqnarray}
Here, $\tilde{\alpha}_{i} = g_{i}^{2}/16\pi^{2}$ and 
$C_{i}^f$ is the
Casimir for $f$, in SU(5) 
normalization, and $Y_{f}$ is its
hypercharge.
The two-loop contribution 
is decoupled at the scale $\mu^{\prime} \sim \widetilde{m}_{1,2}$ of the
heavy scalars, 
which we approximate as  $\sim 10$ TeV \cite{foot4}.
The positive gaugino mass 
contribution, however, survives down to the scale where the gauginos decouple,
which we approximate as $\mu \sim 1$ TeV.  The RGEs can be
solved analytically and the solutions are given by
\begin{eqnarray}
\lefteqn{
	m_{\tilde{f}}^{2} (t)  =  m_{\tilde{f}}^{2} (0)
		+ \sum_{i} \frac{2}{b_{i}} (M_{0}^{2} - M_i^2(t))
						C_{i}^f }
						\nonumber \\
	 & & - 8 \widetilde{m}_{1,2}^{2} \left[
			\left(\frac{1}{2} N_{5} + \frac{3}{2} N_{10} \right)
			\sum_{i} \frac{1}{2 b_{i}}
			(\tilde{\alpha}_{GUT} - \tilde{\alpha}_{i} (t'))
			C_{i}^f \right. \nonumber \\
	 &  & \left.  
			- (N_{5} - N_{10})
			\frac{3}{5} Y_{f}
			\left( \frac{4}{3} 
			\frac{\tilde{\alpha}_{GUT}}{b_{1}-b_{3}}
			\frac{1}{2} \ln
			\frac{\tilde{\alpha}_{1}(t')}{\tilde{\alpha}_{3}(t')}
			\right. \right. \nonumber \\
	& & \left. \left.
			- \frac{3}{4}
			\frac{\tilde{\alpha}_{GUT}}{b_{1} -b_{2}}
			\frac{1}{2} \ln
			\frac{\tilde{\alpha}_{1}(t')}{\tilde{\alpha}_{2}(t')}
			+ \frac{1}{12} \frac{1}{2b_{1}}
			(\tilde{\alpha}_{GUT} -  \tilde{\alpha}_{1}(t') )
			\right) \right] .
\end{eqnarray}
In the above, the $b_{i}$ stand for the gauge coupling beta-function coefficients,
$t=0$ corresponds to the GUT-scale, $\tilde{\alpha}_{GUT} = \tilde{\alpha}_i
(0)$, and the final scales are at 
$t^{(\prime)} = 
\ln \mu^{(\prime)}/M_{GUT}$.  
The final results can be written down explicitly 
in terms of the universal gaugino mass $M_{0} = M_{i}(0)$, the heavy 
scalar mass $\widetilde{m}_{1,2}^{2}$ and the boundary value of the third generation
scalar mass $m_{\tilde{f}} (0)$ as
\begin{eqnarray}
\lefteqn{
m^2_{\tilde{f}}(t) = m^2_{\tilde{f}}(0) + 
\left(.245 C_1^f + .599 C_2^f + 3.20 C_3^f\right) M_0^2} \nonumber\\ 
&&-10^{-2}(.157 C_1^f + .292 C_2^f + .750 C_3^f - .097 Y_f)
N_{10} \widetilde{m}_{1,2}^{2}\nonumber\\   
&&-10^{-2}(.052 C_1^f + .097 C_2^f + .250 C_3^f + .097 Y_f)N_5
\widetilde{m}_{1,2}^{2}
\end{eqnarray}
with $\tilde{\alpha}^{-1}_{GUT} = 25\times 4\pi$.
The combinations of $m_{\tilde{f}}(0)/\widetilde{m}_{1,2}$ and 
$M_{0}/\widetilde{m}_{1,2}$ 
giving vanishing $m_{\tilde{f}}(t)$ for each $f$ are plotted in 
Fig.~\ref{both} for the
case with $N_5 = N_{10} = 2$. The regions below the curves are all excluded.  

\begin{figure}
\centerline{ \epsfxsize = 0.45\textwidth \epsfbox{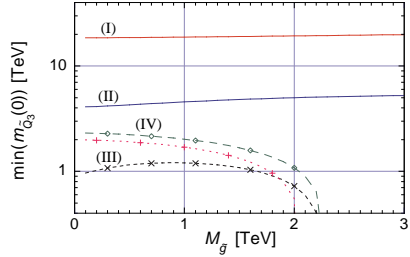} } 
\caption[m3min]{The minimum boundary mass of the left-handed scalar top
$\mbox{min}(m_{\tilde{Q}_3}(0))$ required to avoid
negative $m_{\tilde{Q}_3}^2$ at the weak scale,
while keeping the $\mbox{min}(\widetilde{m}_{1,2})$ 
within the constraints from Fig.~\ref{m1min}.  As in 
Fig.~\ref{m1min}, four cases are considered: (I) 
$(\delta_{12}^{d})_{LL} = (\delta_{12}^{d})_{RR} = 1$, (II) 
$(\delta_{12}^{d})_{LL} = (\delta_{12}^{d})_{RR} = 0.22$, (III) 
$(\delta_{12}^{d})_{LL} = (\delta_{12}^{d})_{RR} = 0.05$, and (IV) 
$(\delta_{12}^{d})_{LL}=0.22$ and $(\delta_{12}^{d})_{RR} =0$.  Two 
curves are shown for the last case.  The upper curve is for $N_{5} = 
N_{10} = 2$ as in other cases, and the constraint is slightly weaker if 
$N_{5} = 0$.}  
\label{m3min}
\end{figure}

Combining this plot with the $\Delta m_K$ constraints, we
obtain lower bounds on $m_{\tilde{Q}_3} (0)$ so that 
$(\Delta m_K)_{\tilde{q},\tilde{g}}$ is on the 
experimental bound while retaining positivity of $m_{\tilde{Q}_3} (t)$. 
The results are shown in Fig.~\ref{m3min} for each pattern of
squark masses (I--IV).
For the cases (I) and (II), 
$m_{\tilde{Q}_3} (0)$ must be at least 
larger than 4~TeV, which is clearly beyond 
whatever can be regarded as natural. For instance, the fine-tuning 
in EWSB quantified in \cite{BG} 
scales as $10 \% \times (m_{\tilde{Q}_3} (0)/300~\mbox{GeV})^{-2}$,
$m_{\tilde{Q}_3} (0) > 4$~TeV requires a severe fine tuning in 
EWSB worse than the $10^{-3}$ level. 
It is clear that one needs further 
alignment or degeneracy to keep $m_{\tilde{Q}_3} (0)$ within a natural 
range.  Case (III), where $(\delta^{d}_{12})_{LL} = 
(\delta^{d}_{12})_{RR} = 0.05$, marginally allows $m_{\tilde{Q}_3} (0)
\sim M_{\tilde{g}} \sim 1$~TeV.  However this mass range still incurs
a fine-tuning in EWSB
at the $1\%$ level.  Case (IV) is no better than this.  In this case 
$(\delta^{d}_{12})_{RR} = 0$, and there is an option to keep the {\bf 
5}$^{*}$ fields of first- and second-generations at the weak scale.  
Fig.~\ref{m3min} shows two curves for this case depending on $N_{5} = 
N_{10} = 2$ as in the other cases or $N_{5} = 0$, $N_{10} = 2$ which 
gives the most conservative constraint.   None of the patterns for scalar 
mass matrices we considered allow $m_{\tilde{Q}_3} (0)$ in the most 
natural range $\sim 100$~GeV.  Recall that the actual constraint is 
stronger than what we presented; we ignored CP-violation in the 
$K^{0}$--$\overline{K}^{0}$ mixing and the top Yukawa 
coupling in the RGE.  We conclude that pushing up the first- 
and second-generation scalar masses does not solve the 
flavor problem, and hence either a relatively 
strong flavor symmetry or a dynamical mechanism to generate 
degenerate scalar masses is necessary.

Finally, we would like to comment on the anomalous U(1) models
\cite{BD,DP,RM,NW,Zhang} 
which naturally generate a split 
mass spectrum between scalars of different generations \cite{foot5}.
These models do not fall into any of the 
patterns (I--IV) we discussed, and hence require a separate 
discussion. The model in \cite{DP} suppresses 
$(\Delta m_{K})_{\tilde{q},\tilde{g}}$ by assigning the same anomalous 
U(1) charges to the first- and second-generations \cite{foot6}, 
thereby making them highly 
degenerate.  However, it predicts a mass spectrum with 
$m_{\tilde{f}} (0) /\widetilde{m}_{1,2} = 0.1$ and 
$M_0/\widetilde{m}_{1,2} =  
0.01$, which is clearly excluded by Fig.~\ref{both}, because
$m^2_{\tilde{Q}_{3}}$ is driven negative.  In \cite{RM}, a
similar choice of 
anomalous U(1) charges is made, with $(
\delta^d_{12})_{LL,RR} \sim m_c/m_t \lesssim .01$ 
It was claimed that
the flavor problem (including the constraint from $\epsilon_K$ allowing 
$O(1)$ CP violating phases) is solved
with the first two generations in the few-TeV range, while keeping the third 
generation and Higgs fields beneath a TeV to achieve natural 
EWSB. However, the constraint from 
$\epsilon_{K}$ used in \cite{RM} was too weak.
For $(\delta^d_{12})_{LL,RR} \sim .01$ and 
$O(1)$ CP violating phases, we find that  
$\widetilde{m}_{1,2}$ must be heavier than 13 TeV, and 
$m_{\tilde{Q}_3} (0)$ must
be heavier than 2.5 TeV in order to avoid being driven negative. 
Ref.~\cite{NW} tries to 
correlate the fermion mass hierarchy to the charges under the 
anomalous U(1), and is hence more realistic.  For instance the
scenario D in \cite{NW} needs 
one {\bf 5}$^{*}$ at 5.0~TeV, another one at 6.1~TeV, and {\bf 
10}$^{*}$ multiplets at 6.1~TeV and 7.0~TeV, respectively, even 
ignoring CP violation.  We obtain 
$m_{\tilde{Q}_3}(0) > 1.0$~TeV, and hence our analysis does not allow $
m_{\tilde{Q}_3}(0)$ in the indicated range of $500~\mbox{GeV}$--1~TeV.
The model is not excluded, but is not better than any of 
the patterns (I--IV) we considered.  If one further implements 
quark-squark-alignment \cite{Zhang}, the situation may be better.  
However, it is then not clear that it is the heavy 
$\widetilde{m}_{1,2}$ which is helping rather than the flavor 
symmetries. 

In summary, we examined the question of whether making first- and 
second-generation scalars heavy can solve the flavor problem 
without relying on flavor symmetries 
or particular dynamical mechanisms to obtain degenerate 
squark masses. In the case where SUSY breaking 
parameters are generated at a high scale, our conclusion is negative.  
Even with an $O(\lambda)$ alignment, 
one needs $\widetilde{m}_{1,2} > 22$~TeV, and 
the contributions to the two-loop RGE of $m^2_{\tilde{Q}_{3}}$
drives it negative unless $m_{\tilde{Q}_3}(0) > 4$~TeV.  Our 
constraints are conservative because we do not include the top Yukawa 
coupling in the RGE and ignored possible CP violation.
A significant degeneracy or much stronger alignment is necessary to 
keep third-generation scalars within their natural range 
$\lesssim (\mbox{a few}\times  100)$~GeV.   

\section*{Acknowledgements}

We thank John March-Russell for collaboration at early stage of this
work.  NAH also thanks Kaustubh Agashe and Michael Graesser for
discussions.  This work was supported in part by  
the U.S. Department of Energy under Contracts DE-AC03-76SF00098 and 
DE-FG-0290ER40542 and in part by the National Science Foundation under 
grant PHY-95-14797.  NAH was also supported by NSERC, and HM by the 
Alfred P. Sloan Foundation.


\begin{thebibliography}{99}
\bibitem{foot2} This is of course a natural consequence of a 
field-theoretical grand unification.  Our analysis does not 
depend on the mechanism to make the scalars within the 
SU(5)-multiplets degenerate.  $D_{Y}$ may cancel in 
other ways as well, see \cite{NW} for examples.
\bibitem{foot4} It is not clear 
at what scale $\mu^*$ one should decouple the contributions of the heavy scalars in 
the RGE; an explicit study of the threshold corrections is necessary
to determine this scale. However, $\mu^*$ should lie between 
$m_{\tilde{f}}$ and
$\widetilde{m}_{1,2}$. In order to obtain 
conservative constraints, we decouple the scalars at a 
very high mass scale 10~TeV from the RGE.
\bibitem{foot5} This class of models predicts small gaugino masses which 
make the situation worse.
\end{thebibliography}
\end{document}